\documentclass{PoS}
\usepackage{graphicx}
\usepackage{amsmath}
\usepackage{dcolumn}
\usepackage{epsfig}
\usepackage{psfrag}
\usepackage{lineno}

\title{Belle II studies of missing energy decays and searches for dark photon production}

\ShortTitle{Belle II studies of missing energy decays and searches for dark photon production}

\author{\speaker{Gianluca Inguglia}%
         \thanks{For the Belle II Collaboration.}\\
        DESY\\
        E-mail: \email{gianluca.inguglia@desy.de}}

\abstract{The Belle II experiment at the SuperKEKB collider is a major upgrade of the KEK ``$B$ factory'' facility in Tsukuba, Japan. The machine is designed for an instantaneous luminosity of $8\times 10^{35}$~cm$^{-2}$\,s$^{-1}$, and the experiment is expected to accumulate a data sample of about 50 ab$^{-1}$ well within the next decade. With this amount of data, decays sensitive to physics beyond the Standard Model can be studied with unprecedented precision. One promising set of modes are physics processes with missing energy such as $B^+\to\tau^+\nu_{\tau}$, $B\to D^{(*)}\tau\nu_{\tau}$, and $B\to K^{(*)}\nu\bar\nu$ decays. The Belle II data also allows searches for candidates for the dark photon, the gauge mediator of a hypothetical dark sector, which has received much attention in the context of dark matter models.}

\FullConference{XXIV International Workshop on Deep-Inelastic Scattering and Related Subjects\\
                 11-15 April, 2016\\
                 DESY Hamburg, Germany}

\begin{document}

\section{Introduction}
The existence of new physics (NP) beyond the standard model (SM) can be probed in the search of its effects in those decays that in the SM are rare and suppressed. In fact the presence of new heavy particles and forces would affect the decay amplitudes and branching ratios (BR) of known mesons, such as the $B$ meson, through an interference term that can be obtained by replacing the SM mediator of a given interaction in a process (for example the $W^+$ boson in $B^+ \to \tau \nu_{\tau}$) with a new physics mediator (for example a charged Higgs boson, $H^+$) in the same process; the existence of such a process would either enhance or further suppress the BR with respect to SM expectations. The Belle II detector at the Super-KEKB accelarator complex will collect data from asymmetric $e^+e^-$ collisions at centre-of-mass (CM) energies equivalent to those of some $\Upsilon(nS)$ resonances and it is therefore perfectly suited for the search of rare $B$ decays. Dark matter (DM) and dark forces can also be searched for at the Belle II Experiment. The lack of experimental evidence to date for the neutralino ($\chi^0$, the most suitable DM candidate in SUSY models with R-parity) has given new light to a class of alternative models and theories in which DM is described as belonging to a new \textit{dark sector} and interacting (in the minimal model) via the force mediator of an extra $U(1)'$ symmetry, the dark photon $A'$, under which only DM particles, neutral under the SM, are charged. If the dark photon is light (with a mass in the GeV/$c^2$ scale) it mixes kinetically with the SM photon making possible the search at the Belle II Experiment. In the next sections the expected Belle II sensitivities to\footnote{Charge conjugate decays will be implied throughout the rest of this manuscript.} $B^+ \to l^+\nu_l$, $B \to D^* \tau^+ \nu_\tau$, $B^0 \to K^* \nu \bar\nu$ will be presented together with discovery potential of the dark photon in the processes $A' \to l^+l^-$ ($l=e,\mu$), and $A' \to \chi\bar\chi$.
 
\section{Belle II sensitivity to new physics in \boldmath{$B^+ \to l^+ \nu_l$} }
The leptonic decays $B^+ \to l^+ \nu_l$ ($l=\mu, \tau$) are tree level decays in which the $\bar b$ and $u$ quarks contained in the $B^+$ annihilate  into a virtual $W^+$ boson involving the $V_{ub}$ elements of CKM matrix~\cite{cabibbo} with a subsequent leptonic decay of the $W^+$, with SM rate given by:
\begin{equation}\label{equation0}
\Gamma(B^{+} \to l^{+} \nu_l)= \frac{G_F^2 m_B m_l^2f_B^2}{8\pi}\vert V_{ub}\vert ^2 (1-\frac{m_l^2}{m_B^2})\tau_B
\end{equation}
where $G_F$ is the Fermi constant, $m_B$ and $m_l$ the masses of the $B$ meson and of the $l$ lepton, $f_B$ the $B$ meson decay constant and $\tau_B$ the $B$ meson lifetime. Since $m_e<m_\mu<m_\tau$ and Eq.~\ref{equation0} contains the suppression factor $m_l^2$, then $\Gamma(B^{+} \to e^{+} \nu_e)<<\Gamma(B^{+} \to \mu^{+} \nu_\mu)<<\Gamma(B^{+} \to \tau^{+} \nu_\tau)$. NP can enhance these decay rates, for example in the supersymmetric two Higgs doublets models (2HDM) the rate can be written as:
\begin{equation}
\Gamma(B^{+} \to l^{+} \nu_l)= \frac{G_F^2 m_B m_l^2f_B^2}{8\pi}\vert V_{ub}\vert ^2 (1-\frac{m_l^2}{m_B^2})\times (1- \frac{\tan^2\beta}{1+\tilde{ \varepsilon_0} \tan\beta} \frac{m_B^2}{m_{H^\pm}^2})^2
\end{equation}
where $\tan\beta$ defines ratio of the vacuum expectation values (\textit{vev}) for two Higgs doublets and $m_{H^\pm}$ is the mass of the charged Higgs bosons. The search for $B^+ \to \tau^+\nu_\tau$ events is (will be) performed with data collected with the Belle (Belle II) detector at the (Super-)KEKB collider operating at the mass of the $\Upsilon(4S)$ which is known to decay to $B^+B^-$ pairs with $BR(\Upsilon(4S)\to B^+B^-)=0.514\pm0.006$ and it relies on the capabilities of software and detector to (fully) reconstruct one $B$ meson ($B_{rec}$) to infer the flavour of and to study the other ($B_{sig}$). The reconstruction of the decays and the search for $B^+\to\tau^+\nu_\tau$ is challenging in the sense that on the one hand one has to apply complex algorithms to reconstruct the $B_{rec}$ affecting the reconstruction efficiency and on the other hand the $BB^+\to\tau^+\nu_\tau$ events are characterised by the presence of large missing energy due to the presence of neutrinos in the final state (respectively one or two neutrinos in hadronic or leptonic $\tau$ decays). Being $E_B$, $p_B$ the energy and momentum of the reconstructed $B$ meson and $E_{beam}$ the beam energy in CM system, the $B_{rec}$ candidates are selected using the standard variables $M_{bc}=\sqrt{E_{beam}^2-P_B^2}$ and $\Delta E=E_B-E_{beam}$. After the reconstruction of $B_{rec}$ candidates, charged tracks are selected to identify the $\tau$ lepton for example in the following decays: $\tau^+ \to \mu^+\nu\bar\nu$, $e^+\nu\bar\nu$, $\pi^+\nu$ ,$\pi^+\pi^0\nu$, $\pi^+\pi^+\pi^-\nu$. The signature of the $B\to\tau\nu$ decay is then searched as an excess of events in the $E_{ECL}$ ($=E_{tot}-E_{B_{rec}}-E_{tracks}$) distribution defining the energy deposition in the electromagnetic calorimeter and that is peaked around zero for signal events. A slight excess of events has been observed by the Belle Experiment in two independent analyses using hadronic and semileptonic decays of the $B_{rec}$, and due to limited statistics upper limits have been set to $BR_{B\to\tau\nu}<(1.79_{-0.49}^{+0.56}(stat)_{-0.51}^{+0.46}(syst))$ for hadronic tagged $B_{rec}$~\cite{hadronic} and to $BR_{B\to\tau\nu}<(1.25\pm0.28\pm0.27)\times 10^{-4}$  and $BR_{B\to\tau\nu}<(1.54\pm0.38\pm0.37)\times 10^{-4}$ for semileptonic tagged $B_{rec}$~\cite{semilep} at $90 \%$ confidence level (C.L.). With 50 $ab^{-1}$ of data that will be collected by the Belle II experiment within the next years and with improved detector and software, one would expect to be able to further constrain this decay to $BR_{B^+ \to \tau^+ \nu_\tau}<4\times 10^{-5}$ and to observe at 5 $\sigma$ statistical significance the decay $B^+ \to \mu^+ \nu_\mu$ (in this last case a 5 $\sigma$ observation is expected with just some 5 $ab^{-1}$ of data) as shown in Table.~\ref{my-label}.
\begin{table}[]
\centering
\caption{Observed (Belle) and expected (Belle II) precision in the determination of  $BR(B^+ \to l^+ \nu_l)$ for different tagging modes including statistical, systematic, and total uncertainties (in percent).}
\label{my-label}
\begin{tabular}{|c|l|l|l|}
\hline 
\textbf{} Process & Statistical &  Systematic & $Total$ \\ 
					& & (reducible, irreducible) &  \\ \hline \hline 
         $BR(B\to \tau \nu)$ (hadronic tag) &  &  &  \\  
         711 fb$^-1$ & 38.0 & (14.2, 4.4) & 40.8 \\  
         5 ab$^-1$  & 14.4 & (5.4, 4.4) & 15.8 \\ 
         50 ab$^-1$ & 4.6 & (1.6, 4.4) & 6.4 \\ \hline 
         $BR(B\to \tau \nu)$ (semileptonic tag) &  &  &  \\  
         711 fb$^-1$ & 24.8 & (18, $_{-9.6} ^{+6.0}$) & $_{-32.2} ^{+31.2}$ \\  
         5 ab$^-1$  & 8.6 & (6.2, $_{-9.6} ^{+6.0}$) & $_{-14.4} ^{+12.2}$ \\ 
         50 ab$^-1$ & 2.8 & (2.0, $_{-9.6} ^{+6.0}$) & $_{-10.2} ^{+6.8}$ \\ \hline 
         $BR(B\to \mu \nu)$ (untagged) &  &  &  \\  
         253 fb$^-1$ & - & (16.4, 3.0) & $<1.7\times 10^{-6}$ \\  
         5 ab$^-1$   & - & (6.2, 30)  & 5$\sigma$ \\ 
         50 ab$^-1$  &  & (2.0, 3.0) & $\gg5\sigma$ \\ \hline \hline
\end{tabular}
\end{table}
\section{Belle II sensitivity to new physics in \boldmath{$B \to D^* \tau^+ \nu_{\tau}$} }
The semileptonic decay $B \to D^* \tau^+ \nu_{\tau}$ is a $b \to c$ transition proceeding via the emission of a virtual $W^+$ boson in a tree level decay topology. NP can affect this decay in different ways, modifying for example either the BR or the $\tau$ polarization. If NP depends on the mass scale and it is proportional to it as for the case for a charged Higgs boson, in the calculation of branching ratio $B \to D^* \tau^+ \nu_{\tau}$ a term in which the $W^+$ boson is replaced by a charged Higgs boson $H^+$ has to be added, and due to the proportionality to the mass the effect is expected to be more pronounced in decays involving a $\tau$-lepton in the final state with respect to the other two lighter leptons, making such an effect detectable~\cite{2hdm}. An alternative possibility of NP affecting the BR  with respect to the charged Higgs boson is represented by an additional transition (interfering with the SM) in which a virtual leptoquark is produced in the process $b \to \nu_\tau \tilde{h}^*$ and subsequent decay $\tilde{h}^* \to c \tau$~\cite{leptoquark}.\\Two very important quantities that characterise these decay(s) are $R(D)$ and $R(D^*)$ defined as
\begin{equation}\label{eqn:1}
R(D^{(*)})= \frac{\Gamma(B \to D^{(*)} \tau^+ \nu_{\tau})}{\Gamma(B^0 \to D^{(*)} l^+ \nu_{l})_{l=\mu,e}}
\end{equation}
that can be measured experimentally and for which very precise predictions from the SM exist:
\begin{equation}\label{eqn:2}
R(D)=0.297\pm 0.017, 
\end{equation}
\begin{equation}\label{eqn:3}
R(D^*)=0.252\pm 0.003.
\end{equation}
Results reported by the Belle (hadronic tag), $BABAR$ (hadronic tag) and LHCb Collaborations while in agreement with each other have shown a significant deviation from the SM predictions shown in Eqs.~\ref{eqn:2} and \ref{eqn:3} and their averages are $R(D^*)=0.322\pm 0.018\pm 0.012$ and $R(D)=0.391\pm 0.041 \pm 0.028$~\cite{dstar}. The deviation  of the combined results on $R(D)$ and $R(D^*)$ is then found to be 3.9 $\sigma$ from SM prediction. At Belle the strategy for the selection of event candidates proceed via a two steps process. First an algorithm based on the hierarchical reconstruction of the $B_{rec}$ using NeuroBayes is applied, then a check is performed on the remaining particles seen in the detector to evaluate whether these are consistent with signal signature or not. Recently new results from the Belle Collaboration based on semileptonic decays of the $B_{rec}$ have been released in which $R(D^*)=0.302\pm 0.030 \pm 0.011$ showing an improvement over previously published results~\cite{dstarsl}. The new results are compatible with both the SM and the type-II 2HDM for $\tan\beta/m_H \simeq 0.7$ GeV$^{-1}$ (to be compared to $\tan\beta/m_H \simeq 0.5$ GeV$^{-1}$ obtained for the hadronic tag analysis). This decay is also sensitive to the tensor operator in leptoquarks (LQ) models, in particular the $R_2$ LQ model is a good model for compatibility test and assuming $M_{LQ}\simeq 1$ TeV/c$^2$ results show that this model with Wilson coefficient $C_T=+0.36$ is disfavoured. Due to the fact that this study is limited by the efficiencies in the full reconstruction and suffer by dominant systematic effects arising from the limited MC sample used for the defining the PDF shape and from limited knowledge of probability density function (PDF) shapes in $B\to D^{**}l\nu_l$ one can expect that with larger data and MC samples and with improved software at Belle II one will achieve a large improvement in the determination of $R(D)$ and $R(D^*)$ as shown in Table~\ref{my-label-1}.
\begin{table}[]
\centering
\caption{Observed (Belle) and expected (Belle II assuming hadronic tagged $B_{rec}$ mesons.) precision in the determination of $R(D^{(*)})$ including statistical, systematic, and total uncertainties (in percent).}
\label{my-label-1}
\begin{tabular}{|c|l|l|l|}
\hline 
\textbf{} Process & Statistical &  Systematic & $Total$ \\ 
					& & (reducible, irreducible) &  \\ \hline \hline 
         $R(D)$  &  &  &  \\  
         423 fb$^-1$ & 13.1 & (9.1, 3.1) & 16.2 \\  
         5 ab$^-1$  & 3.8 & (2.6, 3.1) & 5.6 \\ 
         50 ab$^-1$ & 1.2 & (0.8, 4.4) & 3.4 \\ \hline 
         $R(D^*)$  &  &  &  \\  
         423 fb$^-1$ & 7.1 & (5.2, 1.9) & 9.0 \\  
         5 ab$^-1$  & 2.1 & (1.5, 1.9) & 3.2 \\ 
         50 ab$^-1$ & 0.7 & (0.5,1.9) & 2.1 \\ \hline 
         \hline
\end{tabular}
\end{table}
\section{Belle II sensitivity to new physics in \boldmath{$B \to K^* \nu \bar\nu$} }
Flavour changing neutral currents (FCNCs) are known to be forbidden in the SM at three level and proceed via loops (penguins or box diagrams) and can be used in the search for new physics. The decays $B^+\to K^{(*)+}\nu \bar\nu$, $B^+\to K^{0}\nu \bar\nu$ and  $B^{0}\to K^{(*)0}\nu \bar\nu$ are $b \to s$ FCNCs that are very suppressed in the SM ($BR[B^+\to K^{*+}\nu\bar\nu]=6.8\pm 2.0\times 10^{-6}$ and  $BR[B^+\to K^+\nu\bar\nu]=4.4\pm 1.5\times 10^{-6}$) and free of uncertainties deriving from long-distant hadronic effects; in addition these transitions can only be searched for in $e^+e^-$ collision making them  \textit{golden modes} to be searched for at the Belle II Experiment. Signal candidates are selected by the full reconstruction of the accompanying $B_{rec}$ meson using the variables $M_{BC}$ and $\Delta E$ and requiring only a $K^{(*)+}$ to be seen as a product of the $B_{sig}$ decay. The signal, as for the decay $B^+ \to l^+ \nu$, consists of an excess of events around zero in the distribution of the variable $E_{ECL}$. Searches for these decays have been performed by the Belle and the $BABAR$ Collaborations where no significant excess of events have been observed and upper limits to the BR have been set to $BR(B^+\to K^+\nu \bar\nu<1.7\times 10^{-5})$ at 90$\%$ C.L. and $BR(B^0\to K^{*0}\nu \bar\nu<1.7\times 10^{-5})$ at 90$\%$ C.L.~\cite{kstarbabar}~\cite{kstarbelle}, these results are well above SM prediction and new physics might still play a role here. In addition to dominant backgrounds coming from $e+e- \to q^+q^-$ continuum events and from $B$ decays involving a $b \to c$ that are suppressed by applying dedicated algorithms and kinematic constrains, the search is challenged by the presence of decay topologies that can mimic the signal, such as $B \to f_2' K*$ followed by $f_2' \to K_L^0 K_L^0$, $B \to \eta_c  K^+$ followed by $\eta_c \to K_L^0 K_L^0$ and $B \to D^0 X$, were $X$ is any meson, followed by $D^0 \to K_L^0 \pi^0$. To reject these backgrounds an efficient $K_L^0$ veto is required. Based on Belle results and on ongoing work on background rejection algorithms at Belle II, a sensitivity to $BR(B^+ \to K^+ \nu \bar\nu,B^0 \to K^0 \nu \bar\nu, B^0 \to K^{*0} \nu \bar\nu)$ at a level of  $0.7-2 \times 10^{-6}$ is foreseen allowing one, for example, to be able to probe SM expectations for $B \to K^{*0} \nu \bar\nu$ at 5$\sigma$ credibility level.
 \section{Belle II discovery potential of the dark photon}
The dark photon $A'$ is the mediator of a hypothetical dark force related to a $U(1)'$ extension of the SM~\cite{darkphoton}. The general idea  is that the dark photon might represent a portal between dark matter particles and standard model particles through the kinetic mixing between the SM photon $\gamma$ and $A'$, $\epsilon F^{Y,\mu\nu}F_{\mu\nu}^D$, in the interaction Lagrangian. Dark matter particles would then be neutral under $SU(3)_C \times SU(2)_L \times U(1)_Y$ and charged under $U(1)'_D$ while standard model particles would be neutral under $U(1)'$ and the $A'-\gamma$ kinetic mixing term would allow $A'$ to decay in SM particles with very small couplings. Due to the expected low mass for $A'$ in the range between few MeV$/$c$^2$ to few GeV$/$c$^2$~\cite{nima}, $A'$ could be produced in $e^+e^-$ collisions at $B$ and $\tau$-charm factories or in dedicated fixed target experiments in process that would depend on its mass and its lifetime. In $e^+e^-$ collisions the dark photon is searched for in the reaction $e^+e^- \to \gamma_{ISR} A'$ with subsequent decays of the dark photon to SM final states $A'\to l^+l^-$,$h^+h^-$ ($l=$leptons, $h=$hadrons) or to dark matter $A'\to \chi \bar\chi$, depending on kinematic constraints. The signatures of its production and decays are characterised either by the presence of an energetic photon in the final state plus two oppositely charged tracks with invariant mass equivalent to that of the dark photon in the case of decays to SM final states or by a mono-energetic photon (for on-shell production), that being $s$ the CM energy of the collision would have an energy $E_\gamma= \frac{s-M_{A'}^2}{2\sqrt{s}}$, plus missing energy in the case of decays to dark matter. For SM decays of $A'$ an additional possibility derives from its lifetime, in fact for a short-lived A' one would expect the dark photon to decay promptly near the interaction region but in case of long-lived A' the decay can happen far from the production point, in which case the two tracks would form a vertex at a significantly displaced position with respect to the interaction region. Searches for $A'$ are ongoing and are challenged by high background levels (prompt decays), by low trigger efficiencies (displaced decays), and by the need of a single photon trigger (decays into dark matter) that was not available at the Belle Experiment (but will be implemented at Belle II). If $A'$ will not be observed with the available Belle data, based on preliminary results of the ongoing searches it is possible to anticipate that dark photon decays to any of the discussed final states with kinetic mixing $\varepsilon\simeq 10^{-4} $ will be well within reach with the full Belle II data sample.


\begin{thebibliography}{99}
\bibitem{cabibbo}
N. Cabibbo, Phys. Rev. Lett. \textbf{10}, 531 (1963);
M. Kobayashi and T. Maskawa, Prog. Theor. Phys. \textbf{49}, 652 (1973).
\bibitem{chargedhiggs}
W. S. Hou, Phys. Rev. D \textbf{48}, 2342 (1993); S. Baek and Y. G. Kim, Phys. Rev. D \textbf{60}, 077701 (1999); H. Baer et al., Phys. Rev. D \textbf{85}, 075010 (2012).
\bibitem{hadronic}
Belle Collaboration, Phys. Rev. Lett. \textbf{97}, 251802 (2006).
\bibitem{semilep}
Belle Collaboration, Phys. Rev. D \textbf{92}, 051102(R) (2015); Belle Collaboration, Phys. Rev. D \textbf{82}, 071101(R).
\bibitem{2hdm}
Minoru Tanaka and Ryoutaro Watanabe, Phys. Rev. D \textbf{87}, 034028.
\bibitem{leptoquark}
Y. Sakaki, R. Watanabe, M. Tanaka, and A. Tayduganov, Phys. Rev. D \textbf{88}, 094012 (2013).
\bibitem{dstar}
Belle Collaboration, Phys. Rev. Lett. \textbf{99}, 191807 (2007); Belle Collaboration, Phys. Rev. D \textbf{82},
072005 (2010); Belle Collaboration, Phys. Rev. D \textbf{92}, 072014 (2015); $BABAR$ Collaboration, Phys. Rev. Lett.
\textbf{109}, 101802 (2012); LHCb Collaboration, Phys. Rev. Lett. \textbf{115}, 111803 (2015).
\bibitem{dstarsl}
Belle Collaboration, Belle-CONF-1602, arXiv:1603.06711v1.
\bibitem{kstarbabar}
$BABAR$ Collaboration, Phys. Rev. D \textbf{87}, 112005.
\bibitem{kstarbelle}
Belle Collaboration, Phys. Rev. D \textbf{87}, 111103(R).
\bibitem{darkphoton}
P. Fayet, Phys. Lett. B \textbf{95}, 285 (1980), P. Fayet Nucl. Phys. B \textbf{187}, 184 (1981).
\bibitem{nima}
N. Arkani-Hamed et al. Phys. Rev. D \textbf{79}, 015014, 2009.
\end{thebibliography}
\end{document}